\documentclass[aps, prd]{revtex4-1} 
\usepackage{fancyhdr}
\usepackage[margin=2.4cm]{geometry}
\usepackage{tensor}
\usepackage{amsmath}
\usepackage{amssymb}
\usepackage{amsthm}
\usepackage{bbm}
\usepackage{physics}
\usepackage{slashed}
\usepackage{blindtext}
\usepackage{graphicx}
\usepackage{url}

\begin{document}
\title{Hadamard Renormalization of a Two-Dimensional Dirac Field}
\author{Adam G. M. Lewis}
\affiliation{Perimeter Institute for Theoretical Physics, 31 Caroline St. N., Waterloo, Ontario, Canada, N2L 2Y5}
\email{alewis@perimeterinstitute.ca}

\date{\today}

\begin{abstract}
The Hadamard renormalization procedure is applied to a free, massive Dirac field $\psi$ on a two-dimensional Lorentzian spacetime. This yields the state-independent
divergent terms in the Hadamard bispinor $G^{(1)}(x, x') = \frac{1}{2} \left\langle \left[ \bar{\psi}(x'), \psi(x) \right] \right\rangle$ as $x$ and $x'$ are brought together
along the unique geodesic connecting them. Subtracting these divergent terms within the limit assigns $G^{(1)}(x, x')$, and thus any operator expressed in terms of it, a finite value at the coincident point
$x' = x$. In this limit, one obtains a quadratic operator instead of a bispinor. The procedure is thus used to assign finite values to various quadratic operators, including the stress-energy tensor. Results are presented covariantly, in a conformally-flat coordinate chart at purely spatial separations, and in the Minkowski metric. These terms can be directly subtracted from combinations of $G^{(1)}(x, x')$ - themselves obtained, for example, from a numerical simulation - to obtain finite expectation values defined in the continuum. 
\end{abstract}

\maketitle
\section{Motivation}
Considering that one's surroundings tend to at least seem four-dimensional, two-dimensional models play a surprisingly central role in quantum field theory. This is largely a matter of convenience: they tend to be more readily soluble by non-perturbative methods than their higher-dimensional counterparts. The Thirring model \cite{THIRRING} of self-interacting two-dimensional fermions, for example, can be solved non-perturbatively in the massless \cite{Johnson1961, Hagen1967} and massive cases \cite{osti_4825853}.  Enough results have by now been gathered about two-dimensional \emph{conformal} field theories \cite{BELAVIN1984} to form practically a new discipline.

Such non-perturbative studies often (e.g. \cite{PhysRevB.14.2153}) exploit an identification between two-component Dirac spinors and continuum limits of spin chains. Here, Dirac spinors in the continuum are mapped, for example by ``staggering'' \cite{KogutSusskindLattice, Susskind_Lattice_1976, SusskindLattice1976}, to continuum limits of fermionic lattice operators. These, in turn, can be exactly related to Pauli operators by a so-called Jordan-Wigner transformation \cite{JordanWigner}. 

This strategy has enjoyed something of a renaissance of late, because the numerical simulation of spin-chain Hamiltonians using ``matrix product state'' \cite{AffleckDMRG, Hinrichsen_1996, Klumper_1993, Fannes_1992, PhysRevLett.93.227205, VidalMPS1, VidalMPS2, VidalMPS3} algorithms has grown sufficiently mature for application to lattice field theory \cite{bauls2019phase, bauls2017tensor, PhysRevD.66.013002,Sugihara_2005,PhysRevX.4.041024,PhysRevLett.113.091601,PhysRevLett.112.201601,Banuls:2013zva,PhysRevX.7.041046,CICHY20131666,PhysRevX.6.011023,PhysRevD.96.114501,PhysRevD.93.094512,PhysRevD.94.085018,PhysRevD.92.034519}.
Such studies encounter a recurrent difficulty during computations of certain ``quadratic'' expectation values. These diverge in the continuum limit, and some sort of counter-term must be subtracted to get finite results. Most of the standard renormalization techniques used in perturbation theory are not available in a numerical lattice setting, since the ``bare'' results are found directly in coordinate space, and since the forms of divergences with the lattice spacing are not easily determined.

A similar problem occurs during studies of quantum field theory in curved spacetime \cite{BirrellAndDavies, ParkerAndToms}. On general manifolds, one has neither a preferred vacuum state nor spatial homogeneity, so that one must in this case as well regulate and renormalize directly in coordinate space. Among the various techniques for doing this, covariant point-splitting followed by Hadamard renormalization \cite{hadamard1910lecons, ChristensenReg1978, 1984Ottewill, DEWITT1960, ChristensenVev1976, DecaniniFolacci, AMBRUS2015} is particularly interesting for application to numerics \cite{lewis2019classical}, since it works by subtracting pre-computed terms from externally-supplied two-point functions, which are relatively amenable to simulation.

Curved two-dimensional spacetimes are of some physical interest, despite the identical vanishing of the Einstein tensor upon them. Most of the familiar QFT-in-curved-spacetime effects such as Hawking radiation \cite{HawkingRad1974} occur in both two \cite{PhysRevD.14.870} and four dimensions, but the former are much simpler technically. They also arise theoretically from ``dilaton'' theories \cite{PhysRevD.45.R1005, PhysRevD.47.2454, GRUMILLER2002327}, obtained, for example, by restricting Einstein gravity to spherical symmetry and then compactifying. Even in flat spacetime, Hadamard renormalization is of some use. Though the actual flat-spacetime divergences can be more easily calculated, the Hadamard procedure establishes their independence of the particular quantum state.

Once the terms have been computed, renormalization is as simple as subtracting them from correlation functions, but unfortunately the initial computation is a bit involved. Early studies found results in 4 dimensions, first for scalar, and then for vector and spinor fields. More recent work has concerned a scalar field in dimensions from 2 up to 6 \cite{DecaniniFolacci}. In the special case of AdS spacetime, where parallel transport operators can be found in closed form \cite{AMBRUSCQG2017}, Hadamard renormalization has been used to obtain exact expectation values, both for vacuum states of scalars \cite{KENT2015} and spinors \cite{AMBRUS2015}, and for thermal states of both \cite{AMBRUSAIP2017, AMBRUS2018}. The close-distance singularities int he two-point functions of chiral fermions one arbitrary two-dimensional backgrounds has also been studied \cite{Farhang2017, Farhang2018}.

The immediately relevant case of a Dirac field in two dimensions, though, seems to be missing. This is the most useful case for application to matrix product state simulations, whose expense scales with the spatial dimension and with the local Hilbert space of the lattice model. I develop the missing results here.

Section II establishes notation, while Section III reviews the well-understood theory of two-point functions on curved spacetime. Section IV details the replacement of quadratic
expectation values with divergent limits of correlation functions. Section V then shows how to compute the divergent terms in the correlation functions, and Section VI how to relate the latter to those regulating the quadratic operators. Section VII then specializes these results to a general conformally flat metric at zero coordinate time separation, and to Minkowski spacetime.

As an accessory to the main text, Mathematica \cite{Mathematica} notebooks that adapt the package xTensor \cite{NUTMA20141719, MARTINGARCIA2007640, MARTINGARCIA2008586, MARTINGARCIA2008597} to the manipulation of bispinors and to their coincidence limits are hosted at \cite{GitHubNotebooks}. These are not restricted to two dimensions. One of them, in particular, performs the computation of the divergences in the stress-energy tensor from those in the Hadamard function, which otherwise is quite involved. 

\section{The Free Dirac Field}
Consider a globally hyperbolic, two-dimensional spacetime $\mathcal{M}$ with metric $g_{\mu \nu}(x)$ and Lorentzian signature. Each point $x \in \mathcal{M}$ will be equipped with a spinor field $\psi_A (x)$, whose two components are addressed by uppercase Latin indices $A$. Spinor indices will, however, normally be suppressed, $\psi_A(x) \mapsto \psi(x)$. The spinor fields are defined to transform under a local Lorentz transformation like (see the discussion surrounding (5.256) of \cite{ParkerAndToms})
\begin{equation}
\label{eq:SpinorDef}
\psi_A (x) \mapsto \psi'_A = S(L(x)) \psi_A (x)
\end{equation}
where $S(L(x))$ is some representation of a double covering of the Lorentz group. 
To relate between local Lorentz transformations and global diffeomorphisms, introduce a set of local frame fields $e\indices{^a_\mu}(x)$ via
\begin{subequations}
\label{eq:FrameFields}
\begin{align}
e\indices{^a_\mu}(x) e\indices{^b_\nu}(x) \eta_{ab} &= g_{\mu \nu}(x), \\
e\indices{_a^\mu} (x) e\indices{^a_\nu} (x) &= \delta^\mu_\nu, \\
e\indices{_a^\mu}(x) e\indices{^b_\mu}(x) &= \delta^a_b,
\end{align}
\end{subequations}
where $\delta^\mu_\nu$ is the Kronecker delta. The lowercase-Latin ``Lorentz'' indices are raised and lowered by the Minkowski metric $\eta_{a b} = \mathrm{diag} \begin{pmatrix} -1 & 1 \end{pmatrix}$, just as the Greek ``world" indices are by the spacetime metric $g_{\mu \nu}(x)$. Relationships between uppercase-Latin and, respectively, lowercase-Latin and Greek indices, are furnished, respectively, by the ``flat spacetime" gamma matrices $\gamma\indices{_{aB}^A}$ and ``curved spacetime" gamma matrices $\tilde{\gamma}\indices{_{\mu B}^A}(x)$, themselves defined by their anticommutation relations
\begin{align}
\{\gamma^a, \gamma^b \} &= 2 \eta^{a b}, \\
\{ \tilde{\gamma}^\mu(x), \tilde{\gamma}^\nu(x) \} &= 2 g^{\mu \nu}(x).
\end{align}
The former imply the latter, given
\begin{equation}
\tilde{\gamma}^\mu(x) = e\indices{_a^\mu}(x) \gamma^a.
\end{equation}
where again uppercase Latin spinor indices will usually be suppressed, $\gamma\indices{_{aB}^A} \mapsto \gamma_a$, $\tilde{\gamma}\indices{_{\mu B}^A}(x) \mapsto \tilde{\gamma}_\mu (x)$. 

We will denote covariant derivatives interchangeably with semicolons and nablas (e.g. $\nabla_\mu T_{\nu}(x) = T_{\nu ; \mu}(x)$), and partial derivatives interchangeably with colons and dels (e.g. $\partial_\mu T_{\nu}(x) = T_{\nu, \mu}(x)$). Note the coordinate dependence of the covariant derivative operator will always be suppressed.The covariant derivative of a spinor $\psi_{;\mu}(x)$ is given by
\begin{align}
\label{eq:SpinCovariantDerivative}
\psi_{;\mu}(x) &= \psi_{,\mu}(x) + \zeta_\mu(x) \psi(x), \\
\bar{\psi}_{;\mu}(x) &= \bar{\psi}_{,\mu}(x) - \bar{\psi}(x) \zeta_\mu(x), 
\end{align}
where the spin connection $\zeta_\mu(x)$ stands in for
\begin{align}
\zeta_\mu(x) &= \frac{1}{2} \omega\indices{_{\mu ab}}(x) \Sigma^{ab}, \\
\label{eq:CliffordMatrices}
\Sigma_{ab} &\equiv \frac{1}{4} [\gamma^a, \gamma^b].
\end{align}
Some references call $\omega\indices{_{\mu ab}}(x)$ the spin connection. Whatever its name, it relates to the Christoffel connection $\Gamma\indices{^\lambda_{\mu \nu}}$ via
\begin{equation}
\label{eq:SpinToChristoffel}
\omega\indices{_\mu^a_b}(x) = -e\indices{_b^\nu}(x)( \partial_\mu e\indices{^a_\nu}(x) - \Gamma\indices{^\lambda_{\mu\nu}}(x) e\indices{^a_\lambda}(x) ).
\end{equation}
In the torsion-free geometries we consider, $\Gamma\indices{^\lambda_{\mu \nu}}(x) = \Gamma\indices{^\lambda_{\nu \mu}}(x)$. Given that along with \eqref{eq:SpinToChristoffel} and \eqref{eq:FrameFields}, we find
\begin{equation}
\label{eq:SpinConnectionAntiSym}
\omega\indices{_{\mu a b}}(x) = - \omega\indices{_{\mu b a}}(x).
\end{equation}
 The failure of the spin covariant derivative to commute is measured by
\begin{equation}
\label{eq:SpinorCovariantCommute}
[\nabla_\mu, \nabla_\nu] \psi(x) = \frac{1}{2} \Sigma_{ab} R\indices{^{ab}_{\mu \nu}}(x) \psi(x),
\end{equation}
where $R\indices{^{\mu \nu}_{\sigma \tau}}(x)$ is the Riemann curvature tensor. Partial derivatives commute, of course.

From \eqref{eq:SpinorCovariantCommute} along with the covariant conservation of the gamma matrices,
\begin{equation}
\label{eq:ManyThingsVanish}
\tilde{\gamma}\indices{^\nu_{;\mu}}(x) = 0, 
\end{equation}
the square of the Dirac operator $\tilde{\gamma}^\mu(x) \nabla_\mu$ is found to be
\begin{equation}
\label{eq:GammaToBox}
(\tilde{\gamma}^\mu \nabla_\mu)^2 = \Box + \frac{1}{4} R(x),
\end{equation}
where $\Box \equiv \nabla^\mu \nabla_\mu$ is the covariant wave operator.
 
The field's dynamics will be set by the free Dirac action,
\begin{equation}
\label{eq:DiracActionCurved}
S = -\int \mathrm{d}^{2} x \sqrt{-g(x)} \bar{\psi}(x)\left(\frac{1}{2}\tilde{\gamma}^\mu(x) \overset{\leftrightarrow}{\nabla}_\mu - m\right)\psi(x).
\end{equation}
where $g(x)$ is the metric determinant, the Dirac adjoint $\bar{\psi}(x)$ is defined by
\begin{equation}
\label{eq:DiracAdjoint}
\bar{\psi}(x) \equiv i \psi^\dag(x) \gamma^0,
\end{equation}
and
left-right arrows over a derivative operator indicate, for example,
\begin{equation}
\bar{\psi}(x) \overset{\leftrightarrow}{\nabla}_\mu \psi(x) = \bar{\psi}(x) \psi_{; \mu}(x) - \bar{\psi}_{;\mu}(x) \psi(x).
\end{equation}
Variation of \eqref{eq:DiracActionCurved} yields the Dirac equation and its adjoint,
\begin{subequations}
\label{eq:DiracEquations}
\begin{align}
\tilde{\gamma}^\mu(x) \psi_{;\mu}(x) - m \psi(x) &= 0, \label{eq:DiracEquation} \\
\bar{\psi}_{;\mu}(x) \tilde{\gamma}^\mu(x) + m \psi(x) &= 0. \label{eq:AdjointDiracEquation}
\end{align}
\end{subequations}

The theory can now be quantized by various equivalent means. We use canonical quantization for reference.  Thus we will impose
the canonical anticommutation relations
\begin{subequations}
\label{eq:Canon}
\begin{align}
\left\{\bar{\psi}(t, \mathbf{x}), \pi(t, \mathbf{y}) \right\} &= i\delta(\mathbf{y}-\mathbf{x}), \\
\left\{\psi(t, \mathbf{x}), \psi(t, \mathbf{y})\right\} &= \left\{\pi(t, \mathbf{x}), \pi(t, \mathbf{y})\right\} = 0
\end{align} 
\end{subequations}
where boldface coordinates are purely spatial, while $\delta(\mathbf{x}-\mathbf{y})$ is the Dirac delta distribution. Here
the canonical momentum $\pi(x)$ is defined in terms of the l and the Lagrangian density $\mathcal{L}(x)$ by,
\begin{align}
\label{eq:CanonicalMomentum}
\pi(x) &\equiv \frac{\partial \mathcal{L}(x)}{\partial \psi_{,0}(x)}, \\
S =& -\int \mathcal{L}(x).
\end{align}

Imposition of the canonical anticommutation relations \eqref{eq:Canon} requires $\psi(x)$ to be operator-valued. As a far-from-obvious consequence, classically-unproblematic expressions involving products of fields, for instance $\bar{\psi}(x) \psi(x)$, yield formally infinite expectation values after quantization. A more sophisticated procedure than simple substitution of operators for classical fields is required in these cases.

On the Minkowski metric this problem, in free theories at least, is solved by normal ordering of mode operators. Normal ordering is unsatisfactory on general metrics because it privileges the coordinate of the associated time-ordering operator. Hadamard renormalization is one of a few possible replacements.

The Hadamard scheme runs essentially as follows. A quadratic expectation value with such as $\langle \bar{\psi}(x) \psi(x) \rangle$ is to be computed against a given quantum state. The expression is first regularized by covariant point-splitting \cite{DEWITT1960, ChristensenVev1976, ChristensenReg1978}, which prescribes its replacement by the $x' \to x$ limit of a correlation function such as $\langle \bar{\psi}(x') \psi(x) \rangle$. These correlation functions are assumed to adopt a special ``Hadamard" form, which can be viewed as a restriction of attention to special quantum states called ``Hadamard" states. The Hadamard form provides sufficient information to compute the divergent terms in the $x' \to x$ limit explicitly. By first subtracting these terms and then taking the $x' \to x$ limit, a finite number is obtained. The original quadratic expectation value is finally identified with that number.

\section{Two-Point Functions On Manifolds}
Before explicating the Hadamard procedure in further detail, it is helpful to review some of the well-understood machinery of two-point functions on manifolds.

As is standard, we label indices transforming at $x'$ with a prime. Thus $A_{\mu \nu'}(x, x')$ transforms separately as a vector with respect to coordinate transformations at either the ``base'' point $x$ or the ``field'' point $x'$. Such an object is known as a ``bitensor", with ``bispinors" defined analogously via local Lorentz transformations at $x$ and $x'$. A ``biscalar" is the special case of a bitensor with no indices.

Note that when $x' = x$, the bitensor $A_{\mu \nu'}(x, x')$ transforms, if nonsingular, as a rank-2 tensor at $x$. This case is recurs sufficiently to earn its own notation; thus
\begin{equation}
[B(x, x')] = B(x, x)
\end{equation}
for some smooth bi-spinor-tensor $B(x, x')$. A more generalized notation is useful, however, to enclose the often-interesting case that $B(x, x')$ expands into multiple terms, some of which may be independently singular. To that end, define
\begin{equation}
[B(x, x')] \equiv \lim_{x' \to x} B(x, x')
\end{equation}
where the ``coincidence limit" from $x'$ to $x$ is to be taken along the unique geodesic connecting those points. Square brackets can then be manipulated in the same way as limits.

One is typically interested in using coincidence limits to construct ``covariant expansions'' of tensors (spinors) in terms of bitensors (bispinors). These are morally similar to Taylor series, with the separation measured by 
Synge's world function $\sigma(x, x')$, a biscalar numerically equal to one-half the squared geodesic proper interval between $x'$ and $x$. 

The idiosyncratic, though standard, practice of denoting covariant derivatives of 
$\sigma(x, x')$ without a semicolon will be adopted: $\sigma^\mu(x, x') \equiv \sigma^{;\mu}(x, x')$, for example. In addition, $\sigma$'s coordinate dependence will be suppressed throughout, $\sigma^\mu(x, x') \mapsto \sigma$. The most important facts about $\sigma$ are
\begin{align}
\left[\sigma \right] &= \left[\sigma^\mu \right] = 0, \\
\sigma^\mu \sigma_\mu &= 2 \sigma.
\end{align}
The latter identity, in particular, implies that coincidence limits involving $\sigma^\mu$ scale numerically like $\mathcal{O}(\sigma^{1/2})$.

Outside of Minkowski space, primed indices will generically suffer a nontrivial parallel transport as the coincidence limit is taken. We express this using the spinor parallel propagator, defined by
\begin{subequations}
\label{eq:ParallelPropDefs}
\begin{align}
\label{eq:Jdef}
\mathcal{J}\indices{_B^{A'}_{;\mu}} (x, x') \sigma^\mu &= 0, \\
[\mathcal{J}(x, x')] &= \mathbbm{1}, 
\end{align}
\end{subequations}
where $\mathbbm{1}$ is the identity on spinor indices, and the vector parallel propagator, defined by
\begin{subequations}
\begin{align}
\label{eq:gdef}
g\indices{_{\rho'}^{\nu}_{;\mu}} (x, x') \sigma^\mu(x, x') &= 0, \\
[g\indices{_{\rho'}^{\nu}}(x, x')] &= \delta_\rho^\nu.
\end{align}
\end{subequations}
The identities \eqref{eq:Jdef} and \eqref{eq:gdef} are, respectively, the parallel transport equations for spinors and vectors. Thus, contraction with a parallel propagator implements parallel transport along the coincident geodesic. 

With its spinor indices suppressed, the spin parallel propagator will appear as
\begin{align}
\mathcal{J}_{B}^{A'}(x, x') &\mapsto \mathcal{J}(x, x'), \\
\mathcal{J}_{A'}^{B}(x, x') &\mapsto \mathcal{J}^{-1}(x, x'). 
\end{align}

\section{Covariant Point-Splitting}
The Hadamard procedure will now be applied to four operators: the chiral condensate $C_I(x)$, the axial condensate $C_5(x)$, the $\mu$-current $j_\mu(x)$, and the stress-energy tensor 
\begin{equation}
T_{\mu \nu}(x) \equiv \frac{-2}{\sqrt{-g(x)}}.
\end{equation}
In the classical theory, we have
\begin{subequations}
\label{eq:QuadraticClassical}
\begin{align}
C_I(x) &= m \bar{\psi}(x) \psi(x),  \\
C_5(x) &= \bar{\psi}(x) \gamma^5 (x) \psi(x), \\
j_\mu(x) &= \bar{\psi}(x) \tilde{\gamma}_{\mu}(x) \psi(x), \\
\label{eq:SET}
T_{\mu \nu} &= \frac{1}{4} \bar{\psi}(x)\left(\tilde{\gamma}_\mu(x) \overset{\leftrightarrow}{\nabla}_\nu + \tilde{\gamma}_\nu(x) \overset{\leftrightarrow}{\nabla}_\mu \right) \psi(x).
\end{align}
\end{subequations}
where $\gamma^5 \equiv i \gamma^0 \gamma^1$. In the limit of flat spacetime, the chiral and pseudo-scalar condensate provide measures of symmetry breaking, while the currents form a conserved quantity. The stress-energy tensor both measures how the field responds to diffeomorphisms and, in general relativity, sources the gravitational field. Unfortunately, the expressions \eqref{eq:QuadraticClassical} become problematic after quantization, since their right hand sides will typically be formally infinite. 

Hadamard renormalization is one of various routes to well-defined generalizations. The first step is to construct ``point-split" correlation functions, depending separately on the ``base" point $x$ and a new ``field" point $x'$, such that the classical expressions \eqref{eq:SET} and \eqref{eq:QuadraticClassical} are defined in the coincidence limit from $x'$ to $x$ along their unique connecting geodesic.  

To make eventual contact with the Dirac equation, it is convenient to express the point-split expressions in terms of the ``Hadamard'' bispinor \footnote{Many references instead use a Green's function such as the Feynman propagator. Possibly up to an overall factor, the terms one obtains are the same.}
\begin{equation}
\label{eq:HadFunc}
G\indices{^{(1)}_A^{B'}} (x, x') \mapsto G\indices{^{(1)}}(x, x')= \frac{1}{2} \left\langle \left[ \psi_A(x), \bar{\psi}^{B'}(x') \right] \right\rangle.
\end{equation}
The superscript ${}^{(1)}$ is a typical notation used to distinguish the Hadamard bispinor from other two-point functions like the Feynmann propagator, which will not appear here. Due to the comma, the square bracket in \eqref{eq:HadFunc} is the commutator, not the coincidence limit. Note finally that the Hadamard bispinor implicitly depends upon (or, from a different point of view, defines) the quantum state.

The point-split expressions \cite{ChristensenReg1978} we use are
\begin{subequations}
\label{eq:FormalPointSplit}
\begin{align}
C_I(x) &\mapsto \mathbbm{C}_I (x, x') = -m \mathrm{Tr} \mathcal{J} G^{(1)}(x, x'), \\
C_5(x) & \mapsto \mathbbm{C}_5(x, x') = -\mathrm{Tr} \mathcal{J} \gamma^5 G^{(1)}(x, x'), \\
j_\mu(x) &\mapsto \mathbbm{j}_\mu(x, x') = - \mathrm{Tr} \mathcal{J}\tilde{\gamma}_\mu G^{(1)}(x, x'), \\ 
T_{\mu \nu}(x) &\mapsto \mathbbm{T}_{\mu \nu}(x, x')  \nonumber \\ &= \frac{1}{8} \mathrm{Tr} \mathcal{J} \tilde{\gamma}_{(\mu}\left( G^{(1)}_{;\nu)}(x, x') - g\indices{_{\nu}^{\nu'}} G^{(1)}_{;\nu')}(x, x') \right).
\end{align}
\end{subequations}
The traces here are over the suppressed spinor indices. The problem of defining e.g. $\left\langle C_I(x) \right\rangle$ then becomes that of defining the coincidence limits e.g. $\left[ \mathbbm{C}_I(x, x') \right]$. 

We will find that, having restricted attention to so-called ``Hadamard'' quantum states, all the divergences in $[G^{(1)}(x, x')]$, and thus in each of \eqref{eq:FormalPointSplit}, are determined only by the Lagrangian and by the geometry local to $x$ and $x'$. We denote by an overbar the terms in a two-point function which are a) fully determined by the Lagrangian and the spacetime geometry, and b) non-vanishing at coincidence. By the above requirements, the coincidence limit of the difference between a two-point function and its barred correspondent is well-defined.

Later, we will develop machinery to compute $\bar{G}^{(1)}(x, x')$ explicitly. Having done so, we can find the locally-determed terms of each point-split two-point function via
\begin{subequations}
\label{eq:DivergentTerms}
\begin{align}
\bar{\mathbbm{C}}_I (x, x') &\equiv -m \mathrm{Tr} \mathcal{J} \bar{G}^{(1)}(x, x'), \\
\bar{\mathbbm{C}}_5(x, x') &\equiv -\mathrm{Tr} \mathcal{J} \gamma^5 \bar{G}^{(1)}(x, x'), \\
\bar{\mathbbm{j}}_\mu(x, x') &\equiv - \mathrm{Tr} \mathcal{J}\tilde{\gamma}_\mu \bar{G}^{(1)}(x, x'), \\ 
\bar{\mathbbm{T}}_{\mu \nu}(x, x')  &\equiv= \frac{1}{8} \mathrm{Tr} \mathcal{J} \tilde{\gamma}_{(\mu}\left( \bar{G}^{(1)}_{;\nu)}(x, x') - g\indices{_{\nu}^{\nu'}} \bar{G}^{(1)}_{;\nu')}(x, x') \right),
\end{align}
\end{subequations}
where it is understood that terms which vanish at coincidence are to be dropped from the right hand side. We can then make the definitions
\begin{subequations}
\label{eq:RenormalizedOperators}
\begin{align}
\langle C_I(x) \rangle &\equiv \left[ \mathbbm{C}_I(x, x') - \bar{\mathbbm{C}}_I(x, x') \right], \\
\langle C_5(x) \rangle &\equiv \left[ \mathbbm{C}_5(x, x') - \bar{\mathbbm{C}}_5(x, x') \right], \\
\langle j_\mu(x) \rangle &\equiv \left[ \mathbbm{j}_\mu(x, x') - \bar{\mathbbm{j}}_\mu(x, x') \right], \\
\langle T_{\mu \nu}(x) \rangle &\equiv \left[ \mathbbm{T}_{\mu \nu} (x, x') - \bar{\mathbbm{T}}_{\mu \nu}(x, x') \right],
\end{align}
\end{subequations}
whose sensibility of course hinges upon the claim that all the coincident divergences within e.g. $\mathbbm{C}_I(x, x')$ are indeed contained within its locally-determined parts e.g. $\bar{\mathbbm{C}}_I(x, x')$. 

\section{Computation of $\bar{G}^{(1)}(x, x')$}
To employ the definitions \eqref{eq:RenormalizedOperators} we must first compute the locally-determined terms $\bar{G}^{(1)}(x, x')$ in the Hadamard function, and next work out those in each point-split operator through \eqref{eq:DivergentTerms}. We will do the first part in this section.

\subsection{The Hadamard Form}
One first assumes that $G^{(1)}(x, x')$ is a homogeneous solution to the Dirac equation
 \begin{equation}
 \label{eq:DiracEquationHadamard}
(\tilde{\gamma}^\mu(x) \nabla_\mu - m)G^{(1)}(x,x') = 0,
\end{equation}
a first order PDE. We would like to appeal to results concerning solutions to second order hyperbolic PDEs. To this end, define the auxiliary propagator $\mathcal{G}(x,x')$ implicitly by
\begin{equation}
\label{eq:AuxPropDef}
(\bar{\gamma}^\mu(x) \nabla_\mu + m)\mathcal{G}(x,x') = G^{(1)}(x, x'). 
\end{equation}
Inserting \eqref{eq:AuxPropDef} into \eqref{eq:DiracEquationHadamard} and applying \eqref{eq:GammaToBox} reveals that $\mathcal{G}(x, x')$ obeys the Klein-Gordon-like equation with a $\zeta = \frac{1}{4}$ curvature coupling,
\begin{equation}
\label{eq:KGeqn}
\left(\nabla^\mu \nabla_\mu + \frac{1}{4} R - m\right)\mathcal{G}(x, x') = 0,
\end{equation}
a second-order hyperbolic PDE. We now make the ``Hadamard" assumption that $\mathcal{G}(x, x')$ is regular except along characteristics of
\eqref{eq:KGeqn}; that is, except when $x$ and $x'$ are lightlike separated. Should this be so it is possible to construct a general form for $\mathcal{G}(x, x')$ even without initial data, called the ``Hadamard form" or ``elementary solution". In two-dimensional spacetime, the Hadamard form is
\begin{equation}
\label{eq:HadForm}
\mathcal{G}(x,x') = \alpha (V(x,x')\ln{\left(\mu \sigma + i \epsilon \right)} + W(x,x'))
\end{equation}
where $V(x,x')$ and $W(x,x')$ are smooth bispinors, and $\epsilon$ a small number that regulates the logarithm. 
The Hadamard form was first presented by Hadamard (pg. 100 of \cite{hadamardCauchy}) as part of a highly recommended 
treatise on the Cauchy problem, apparently the first to draw a distinction between elliptic and hyperbolic PDEs. The relevant results were first imported to relativity theory during a study of the radiation back-reaction by DeWitt and Brehme (\cite{DEWITT1960}), which also laid out much of the necessary theory of bitensors. 

For $\mathcal{G}(x, x')$ of the Hadamard form \eqref{eq:HadForm}, we have
\begin{subequations}
\begin{align}
\label{eq:AuxG}
\bar{\mathcal{G}}(x,x') &= \alpha V(x, x') \ln{\left(\mu \sigma + i \epsilon \right)}, \\ &= \frac{1}{4\pi} V(x, x') \ln{\left(\mu \sigma + i \epsilon \right)},
\end{align}
\end{subequations}
where the overall scale
\begin{equation}
\alpha = \frac{1}{4\pi}
\end{equation}
is fixed in the Appendix by demanding agreement with standard QFT results. In the following, this value will be used for $\alpha$ without comment. The dimensionful ``renormalization parameter" $\mu$ is similarly assigned a value in the Appendix, but we follow tradition in leaving $\mu$ as is during subsequent calculations.

As we will soon discover, the requirement that $\mathcal{G}(x, x')$ be of the form \eqref{eq:HadForm} places a restriction upon the quantum state. States meeting this requirement are said to be ``Hadamard''. Heuristically, such states locally resemble the Minkowski vacuum, and the Hadamard condition is taken \cite{KAY199149, radzikowski1996, Fewster_2013} as a condition for the physical reasonableness of a quantum state. 

The significance of assuming \eqref{eq:HadForm} is perhaps best appreciated by first noting that, using traditional free QFT formalisms involving e.g. mode expansions and normal-ordered Hamiltonian minimization, it is straightforward to construct quantum states failing to meet the Hadamard condition. For example, the ``Rindler" vacuum minimizing the normal-ordered Hamiltonian of constantly-accelerated observers in flat spacetime will fail to meet it, due to additional divergences independent of $\sigma$ at the acceleration horizon.

According to one's taste, Hadamard-renormalized quantum field theory thus either predicts or assumes that standard normal-ordering techniques, generalized to nontrivial metrics, in many cases furnish quantum states which do not occur in nature. Of course free two-dimensional Dirac fields also do not occur in nature, but this qualitative feature of the Hadamard procedure holds more generally.

\subsection{Expansion in $\sigma$}
Due to \eqref{eq:AuxG}, we can determine $\bar{\mathcal{G}}(x, x')$ by computing $V(x, x')$. To do so, insert the Ansatz expansions
\begin{subequations}
\label{eq:VWexpansions}
\begin{align}
V(x, x')  &= \sum_{i=0}^{\infty} V_i(x, x') \sigma^{(i)}, \\
W(x, x') &= \sum_{i=0}^{\infty} W_i(x, x') \sigma^{(i)},
\end{align}
\end{subequations}
into \eqref{eq:HadForm} and then \eqref{eq:KGeqn}. Having done so, the demand that each power of $\sigma$ separately vanish yields a set of recurrence relations for $V_i(x, x')$ and $W_i (x, x')$, along with a boundary condition for $V_0(x, x')$. The resulting expressions are a bit visually confusing, so we will briefly suppress coordinate dependencies while presenting them. They are
\begin{subequations}
\label{eq:RecurrenceRelations}
\begin{equation}
\begin{gathered}
\label{eq:Vrecurrence}
2(n+1)^2 V_{n+1} + 2(n+1) V\indices{_{n+1 ;\mu}} \sigma^{\mu}  \\-2(n+1) V_{n+1} \Delta^{-1/2} \Delta^{1/2}_{;\mu} \sigma^{\mu} \\
+ (\Box_x - m^2 + \frac{1}{4} R) V_n = 0 ,
\end{gathered}
\end{equation}

\begin{equation}
\begin{gathered}
\label{eq:Wrecurrence}
2(n+1)^2 W_{n+1} + 2(n+1) W_{n+1 ;\mu} \sigma^{\mu} \\- 2(n+1) W_{n+1} \Delta^{-1/2} \Delta^{1/2}_{;\mu} \sigma^{\mu}
+4(n+1) V_{n+1} \\+ 2 V_{n+1; \mu} \sigma^{\mu} - V_{n+1} \Delta^{-1/2} \Delta^{1/2}_{;\mu} \sigma^{\mu}
\\+ (\Box_x - m^2 + \frac{1}{4} R) W_n
= 0 ,
\end{gathered}
\end{equation}

\begin{equation}
\label{eq:VBC}
V_{0 ;\mu} \sigma^{\mu} - V_0 \Delta^{-1/2} \Delta^{1/2}_{;\mu} \sigma^{\mu} = 0.
\end{equation}
\end{subequations}
The biscalar $\Delta^{1/2}$, called the van Vleck-Morette determinant, appears here via the identity
\begin{equation}
\label{eq:BoxSigma}
\Box_x \sigma = (d+1) - 2\Delta^{-1/2} \Delta^{1/2}_{;\mu} \sigma^{\mu},
\end{equation}
where $d=1$ is the spatial dimension. It bears repeating that the labels $i$ in $V_i$, e.g. $n+1$ in $V_{n+1}$, are not indices, but simply label the order within the expansions \eqref{eq:VWexpansions}. Note that \eqref{eq:RecurrenceRelations} are formally identical to those obtained in \cite{DecaniniFolacci} for a scalar field with a $\zeta = \frac{1}{4}$ curvature coupling, apart from the differing connection in the covariant derivatives.

In the immediately following subsection we will be able to determine $V_0(x, x')$ from \eqref{eq:VBC} up to a constant scalar coefficient. While the computational effort becomes quickly forbidding, $V_0(x, x')$ in turn provides sufficient information in principle to determine $V(x, x')$ to arbitrary order in $\sigma$ via \eqref{eq:Vrecurrence}. Due to the assumed Hadamard form \eqref{eq:HadForm}, $V_0(x, x')$ also fully determines the divergent terms within each of \eqref{eq:DivergentTerms}. Thus, as promised, these divergent terms are fully determined by the mass and the local spacetime geometry.

On the other hand, $W(x, x')$ additionally depends upon the bispinor $W_0(x, x')$, which is not constrained by any boundary condition. Thus, $W_0(x, x')$ must contain any information apart from the mas ..s and the local spacetime geometry distinguishing different two-point functions from one another. This notably includes the quantum state: all Hadamard states with the same mass and on the same background have the same $\bar{\mathcal{G}}(x,x')$.

\subsection{Solving for $V(x, x')$}
Each time we take a derivative of $V(x, x')$, its scaling with $\sigma$ during the coincidence limit will be reduced by a factor of $\sigma^{1/2}$. Since $\mathbbm{T}_{\mu \nu}(x, x')$ depends on second derivatives of $\mathcal{G}(x, x')$, in order to compute $\bar{\mathbbm{T}}_{\mu \nu}$ we need $V(x, x')$ up to $\mathcal{O}(\sigma)$. In light of \eqref{eq:VWexpansions}, we in turn need $V_0 (x, x')$ up to $\mathcal{O}(\sigma)$, and $V_1(x, x')$ up to $\mathcal{O}(1)$.

To find $V_0(x, x')$ we must solve the boundary equation \eqref{eq:VBC}. To do so, we make the Ansatz 
\begin{equation}
\label{eq:V0Ansatz}
V_0(x, x') = a S(x, x') \mathcal{S}(x, x'),
\end{equation}
where $a$ is a scalar constant, $S(x, x')$ is a biscalar, and $\mathcal{S}(x, x')$ is a bispinor. We then find by inspection that \eqref{eq:VBC}
is satisfied by the simultaneous choices
\begin{subequations}
\begin{align}
\label{eq:Sdef1}
S(x, x') &= \Delta^{1/2}, \\
\label{eq:Sdef2}
\mathcal{S}_{;\mu}(x, x') \sigma^\mu &= 0.
\end{align}
\end{subequations}
A covariant expansion of $\Delta^{1/2}$ can be found, for example, in \cite{ChristensenVev1976}. Up to $\mathcal{O}(\sigma)$ it is
\begin{equation}
\label{eq:DeltaExpansion}
\Delta^{1/2} = 1 + \frac{1}{12} R_{\mu \nu} \sigma^{\mu} \sigma^{\nu} + \mathcal{O}(\sigma^{3/2}), \\ 
\end{equation}
so that $\left[ \Delta^{1/2} \right] = 1$, and
\begin{equation}
\left[ V_0(x, x') \right] = a \left[ \mathcal{S}(x, x') \right].
\end{equation}
In the Appendix we find that, in order to recover the results of standard flat-spacetime QFT, we should demand
\begin{equation}
\left[ V_0(x, x') \right] = -1,
\end{equation}
and thus
\begin{align}
a &= -1, \\
\label{eq:Scoinc}
\left[ \mathcal{S}(x, x') \right] &= \mathbbm{1}.
\end{align}
Taken together, \eqref{eq:Sdef2} and \eqref{eq:Scoinc} form the definition \eqref{eq:ParallelPropDefs} of the spin parallel propagator, so that
$\mathcal{S}(x, x') = \mathcal{J}(x, x')$. We thus have exactly
\begin{equation}
\label{eq:V0result}
V_0(x, x') = -\Delta^{1/2} \mathcal{J}.
\end{equation}

It is not possible to make a covariant expansion of $\mathcal{J}$, but we can insert the expansion \eqref{eq:DeltaExpansion} for $\Delta^{1/2}$ to find
\begin{equation}
\label{eq:V0expansion}
V_0(x, x') = -\mathcal{J} \left( 1 + \frac{1}{12} R \sigma \right) + \mathcal{O}(\sigma^{3/2}),
\end{equation}
where the well-known relations
\begin{subequations}
\label{eq:Riemann2D}
\begin{align}
R_{\mu \nu \sigma \tau} &= \frac{1}{2} R( g_{\mu \nu}(x) g_{\sigma \tau}(x) - g_{\mu \tau}(x) g_{\sigma \nu}(x)), \\
R_{\mu \nu} &= \frac{1}{2} R g_{\mu \nu}(x),
\end{align}
\end{subequations}
which are specific to 2D, were used.

To get $V_1(x, x')$ up to $\mathcal{O}(1)$, we follow \cite{ChristensenVev1976} and first write $\mathcal{J}^{-1} V_1(x, x')$ as a Taylor-like ``covariant'' expansion \footnote{Note that this step, if applied to $\mathcal{J}$, would simply yield $\mathcal{J}^{-1} \mathcal{J} = \mathbbm{1}$, which is why a (useful) covariant expansion of $\mathcal{J}$ is not possible. Similar considerations apply to $\sigma$.},
\begin{equation}
\label{eq:JV1}
\mathcal{J}^{-1} V_1(x, x') = v_0 (x) + v_{1 \mu}(x) \sigma^\mu + \ldots
\end{equation}
where the coefficients $v_i(x)$ depend only on $x$. Taking the coincidence limit of both sides we have
\begin{equation}
\label{eq:v0coin}
v_0 (x) = \left[V_1(x, x')\right].
\end{equation}
Now set $n=0$ in \eqref{eq:Vrecurrence}, insert \eqref{eq:V0expansion}, and take the coincidence limit to find
\begin{equation}
\label{eq:V1coin}
\left[ V_1(x, x') \right] = \frac{1}{24}(-12 m^2 + 5 R) \mathbbm{1},
\end{equation}
\begin{equation}
\label{eq:V1exp}
V_1(x, x') = \mathcal{J} \frac{1}{24}(-12 m^2 + 5 R) + \mathcal{O}(\sigma^{1/2}).
\end{equation}
Combining \eqref{eq:V1exp} and \eqref{eq:V0expansion} with \eqref{eq:VWexpansions}, we find
\begin{equation}
\label{eq:Vexp}
\mathcal{J}^{-1} V(x, x') = -\mathbbm{1} - (1/2)(m^2 +(1/4)R)\sigma\mathbbm{1}.
\end{equation}
Except for the presence of the spinor identity, this is the same result as reported for the scalar field in \cite{DecaniniFolacci} with scalar curvature coupling $\zeta = 1/4$.

\section{The Locally Determined Terms}
We now must insert \eqref{eq:Vexp} into
\eqref{eq:AuxG} and then \eqref{eq:AuxPropDef} to find $\mathcal{\bar{G}}^{(1)}(x, x')$, and then the latter into each of \eqref{eq:FormalPointSplit} to find $\mathbbm{\bar{C}}_{I}(x, x')$, $\bar{\mathbbm{j}}_{\mu}(x, x')$, and $\mathbbm{\bar{T}}_{\mu \nu}(x, x')$.

Doing this, inserting the expansions found in \cite{DEWITT1960, ChristensenVev1976, ChristensenReg1978}, and dropping any terms that vanish at coincidence, one obtains expressions in terms of geometric tensors, $\sigma$, and $\sigma_{\mu}$ only.
This is conceptually straightforward, but a bit tiresome in practice, due to the length of the intermediate expressions involved.
I have written some Mathematica notebooks, based on the package xTensor \cite{NUTMA20141719, MARTINGARCIA2007640, MARTINGARCIA2008586, MARTINGARCIA2008597}, to assist with such bispinor and bitensor manipulations. They can be found at \cite{GitHubNotebooks}. 

Following this procedure yields our central results,
\begin{subequations}
\begin{align}
\label{eq:Condensate}
\mathbbm{\bar{C}}_{I}(x, x') &= -\frac{m^2}{2\pi} \ln{\mu \sigma}, \\
\label{eq:Condensate5}
\mathbbm{\bar{C}}_5(x, x') &= 0, \\
\label{eq:MuCurrent}
\mathbbm{\bar{j}}_\mu(x, x') &= -\frac{\sigma_{\mu}}{2 \pi \sigma}, 
\end{align}
\begin{multline}
\label{eq:Tmunuloc}
\mathbbm{\bar{T}}_{\mu \nu}(x, x') = \frac{1}{4\pi} \left[ \frac{g_{\mu \nu}}{\sigma} - \frac{\sigma_\mu \sigma_\nu}{\sigma^2} + \frac{R}{6}\left(\frac{\sigma_\mu \sigma_\nu}{\sigma} - \frac{5}{4} g_{\mu \nu} \right) \right] \\ + \frac{m^2}{8\pi}\left[\frac{\sigma_\mu \sigma_\nu}{\sigma} + g_{\mu \nu} \left( 1 + \ln{\mu \sigma}\right) \right].
\end{multline}
\end{subequations}

Note that the trace of \eqref{eq:Tmunuloc} is
\begin{equation}
\mathbbm{\bar{T}}\indices{^\mu_\mu}(x, x') = -\frac{R}{48\pi} + \frac{m^2}{2\pi} \left( 1 + \frac{1}{2} \ln{\mu \sigma}\right),
\end{equation}
which, when $m=0$, differs from the standard CFT result \cite{BELAVIN1984} for free Dirac fermions by a factor of $1/2$. If desired, this can be corrected via a procedure due to Moretti \cite{MorettiSET}. Here, the stress-energy tensor is redefined to include a factor proportional to the Lagrangian, which vanishes classically. Thus, define
\begin{equation}
\Theta_{\mu \nu}(x, x') \equiv  g_{\mu \nu}(x) \mathcal{J}^{-1}\left( \tilde{\gamma}^\rho(x) \nabla_\rho - m \right) G^{(1)}(x, x').
\end{equation}
The correction $\Theta_{\mu \nu}(x, x')$ vanishes for the ``classical'' $G^{(1)}(x, x)$ which solves the Dirac equation. However, it does not vanish after the replacement $G^{(1)}(x, x') \to \bar{G}^{(1)}(x, x')$. Instead, 
\begin{equation}
\bar{\Theta}_{\mu \nu}(x, x') = \frac{g_{\mu \nu}}{\pi} \left( \frac{R}{24 \pi} - m^2 g_{\mu \nu} \right).
\end{equation}
Now define
\begin{equation}
\mathbbm{\bar{T}}^{\mathrm{new}}_{\mu \nu}(x, x') \equiv \mathbbm{\bar{T}}_{\mu \nu}(x, x') + q\bar{\Theta}_{\mu \nu}(x, x').
\end{equation}
Then the demand
\begin{equation}
\mathbbm{\bar{T}}\indices{^{\mathrm{new}, \mu}_\mu} = -\frac{R}{24 \pi}
\end{equation}
implies $q=-\frac{1}{4}$, and thus
\begin{multline}
\label{eq:TmunuNew}
\mathbbm{\bar{T}}\indices{^{\mathrm{new}}_{\mu \nu}}(x, x') = \frac{1}{4\pi} \left[ \frac{g_{\mu \nu}}{\sigma} - \frac{\sigma_\mu \sigma_\nu}{\sigma^2} + \frac{R}{2}\left(\frac{\sigma_\mu \sigma_\nu}{3\sigma} - \frac{1}{2} g_{\mu \nu} \right) \right] \\ + \frac{m^2}{8\pi}\left[\frac{\sigma_\mu \sigma_\nu}{\sigma} + g_{\mu \nu} \left( 3 + \ln{\mu \sigma}\right) \right].
\end{multline}
\section{Specialization to Conformally Flat Coordinates}
Our original motivation in computing \eqref{eq:Condensate}, \eqref{eq:MuCurrent}, and \eqref{eq:Tmunuloc} was to regularize numerically-generated data. These will typically be in some specific coordinate chart, localized to equal-time hypersurfaces with $t = t'$. Thus, we outline here how to specialize the given example to a coordinate system, using conformally flat coordinates
\begin{equation}
\label{eq:ConformallyFlat}
g_{\mu \nu}(x) = \Omega^2(x) \eta_{\mu \nu}
\end{equation}
and a purely spatial coordinate separation as a prototype. Such coordinates are always available in 2D.

The connections and the curvature scalar satisfy the component equations
\begin{subequations}
\label{eq:CFgeometry}
\begin{align}
\Gamma\indices{^0_{00}}(x) &= \Gamma\indices{^0_{11}}(x) = \Gamma\indices{^1_{01}}(x) = \frac{\Omega_{,0}(x)}{\Omega(x)}, \\
\Gamma\indices{^0_{01}}(x) &= \Gamma\indices{^1_{00}}(x) = \Gamma\indices{^1_{11}}(x) = \frac{\Omega_{,1}(x)}{\Omega(x)} ,\\
\zeta_0(x) &= \frac{i}{2} \frac{\Omega_{,1}(x)}{\Omega(x)} \gamma^5, \\
\zeta_1(x) &= \frac{i}{2} \frac{\Omega_{,0}(x)}{\Omega(x)} \gamma^5, \\
R &= 2\left( \frac{\Omega_{,1}^2(x) - \Omega_{,0}^2(x)}{\Omega^4 (x)} - \frac{\Omega_{,11}(x)  - \Omega_{,00}(x)}{\Omega^3(x)} \right),
\end{align}
\end{subequations}
while the gamma matrices satisfy
\begin{subequations}
\label{eq:GammaMatrixCF}
\begin{align}
\tilde{\gamma}^\mu(x) &= \Omega^{-1}(x) \gamma^\mu, \\
\tilde{\gamma}\mu(x) &= \Omega(x) \gamma_\mu.
\end{align}
\end{subequations}

Coordinate expansions of $\sigma_{;\mu}$ can be found in \cite{PoissonHad}. Specialized to \eqref{eq:ConformallyFlat} with $t' = t$, they yield
\begin{subequations}
\begin{align}
\label{eq:SigExpansion}
\sigma_0 &=  \frac{1}{2} \Omega(x) \Omega_{,0}(x) r^2 + \mathcal{O}(r^3) \\
\sigma_1 &= -\Omega^2(x) r - \frac{1}{2} \Omega(x) \Omega_{,1}(x) r^2 + \mathcal{O}(r^3)
\end{align}
\end{subequations}
where $r \equiv x' - x$.  We can find expansions of $\sigma$ by inserting these expansions into the identity $\sigma = \frac{1}{2}\sigma^{\mu }\sigma_{\mu}$, and then the result into a Laurent expansion of $\frac{1}{\sigma}$ about $\sigma=0$. The results are 
\begin{subequations}
\label{eq:HadCoordAppendix}
\begin{align}
\mathbbm{\bar{C}}_{I} &= -\frac{m^2}{2\pi} \ln{\frac{1}{2} \mu r^2 \Omega^2(x)}, \\
\mathbbm{\bar{C}}_{5} &= 0, \\
\mathbbm{\bar{j}}_{0} &=\frac{\Omega_{,0}(x)}{2\pi \Omega(x)}, \\
\mathbbm{\bar{j}}_{1} &= -\frac{1}{\pi r} + \frac{\Omega_{,1}(x)}{2\pi \Omega(x)} ,
\end{align}
\end{subequations}
\begin{subequations}
\begin{widetext}
\begin{align}
\mathbbm{\bar{T}}_{00}(x, x') &= \frac{1}{2 \pi r^2} + \frac{1}{2\pi r} \frac{\Omega_{,1}(x)}{\Omega(x)} + \frac{m^2}{8 \pi} \Omega^2(x) \left( 1+ \ln{\mu \frac{1}{2} r^2 \Omega^2(x)}\right) - \frac{5}{96\pi} R \Omega^2 + \frac{1}{24\pi} \frac{\Omega_{,1}^2(x)}{\Omega^2(x)} - \frac{1}{6\pi} \frac{\Omega_{,11}(x)}{\Omega(x)} + \frac{5}{24\pi} \frac{\Omega_{,0}^2(x)}{\Omega^2(x)}, \\
\mathbbm{\bar{T}}_{01}(x, x') &= \frac{1}{2\pi}\left( -\frac{\Omega_{,0}(x)}{\Omega(x)} \frac{1}{r} + \frac{\Omega_{,0}(x)\Omega_{,1}(x)}{2 \Omega^2(x)} - \frac{1}{3} \frac{\Omega_{,01}(x)}{\Omega(x)}\right),  \\
\mathbbm{\bar{T}}_{11}(x, x') &= \frac{1}{2 \pi r^2} + \frac{1}{2\pi r} \frac{\Omega_{,1}(x)}{\Omega(x)} - \frac{m^2}{8 \pi} \Omega^2(x) \left(3 + \ln{\frac{1}{2}\mu r^2 \Omega^2(x)}\right) + \frac{1}{32 \pi} R \Omega^2(x) + \frac{1}{24\pi} \frac{\Omega_{,1}^2(x)}{\Omega^2(x)} - \frac{1}{6\pi} \frac{\Omega_{,11}(x)}{\Omega(x)} + \frac{5}{24\pi} \frac{\Omega_{,0}^2(x)}{\Omega^2(x)},
\end{align}
\end{widetext}
\end{subequations}
\begin{subequations}
\begin{align}
\Theta_{00}(x, x') &= \frac{q}{\pi} \Omega^2(x) \left(m^2 + \frac{R}{24}\right),\\ 
\Theta_{01}(x, x') &= 0,\\
\Theta_{11}(x, x') &= -\Theta_{00}(x, x').
\end{align}
\end{subequations}
Further specializing to Minkowski space, $\Omega=1$, we have
\begin{subequations}
\label{eq:HadMinkowskiAppendix}
\begin{align}
\mathbbm{\bar{C}}_{I}(x, x') &= -\frac{m^2}{2\pi} \ln{\frac{1}{2} \mu r^2}, \\
\mathbbm{\bar{C}}_{5} &= 0,  \\
\mathbbm{\bar{j}}_{0} &= 0, \\
\mathbbm{\bar{j}}_{1}(x, x') &= -\frac{1}{\pi r},
\end{align}
\end{subequations}
\begin{subequations}
\begin{align}
\mathbbm{\bar{T}}_{00}(x, x') &= \frac{1}{2\pi r^2} + \frac{m^2}{8\pi} \left(1 + \ln{\frac{1}{2}\mu r^2} \right),\\
\mathbbm{\bar{T}}_{01} &= 0, \\
\mathbbm{\bar{T}}_{11}(x, x') &= \frac{1}{2\pi r^2} - \frac{m^2}{8\pi} \left(3 + \ln{\frac{1}{2}\mu r^2}\right),
\end{align}
\end{subequations}
\begin{subequations}
\begin{align}
\Theta_{00} &= \frac{q}{\pi} m^2,\\ 
\Theta_{01} &= 0,\\
\Theta_{11} &= - \Theta_{00}.
\end{align}
\end{subequations}

\section{Acknowledgements}
I would like to thank Eric Poisson for extremely useful feedback. Valter Moretti supplied some important clarifications to certain technical points in the derivation. Victor E. Ambru\c{s} provided useful guidance regarding the intricacies of spinor fields. It would have been very difficult to complete this study without the Mathematica package xTensor \cite{NUTMA20141719, MARTINGARCIA2007640, MARTINGARCIA2008586, MARTINGARCIA2008597}, and in particular without guidance in that package's use, provided especially by Leo C. Stein and Jos\'{e} Mart\'{i}n-Garc\'{i}a.

\appendix
\section{Identifications with standard QFT}
In this Appendix we fix the constants $\mu$ and $\alpha$ in \eqref{eq:HadForm} along with $a$ in \eqref{eq:V0Ansatz} by comparison with mode-sum-based QFT in flat spacetime. Choose $g_{\mu \nu}(x) = \eta_{\mu \nu}$ and consider $x'$ and $x$ separated by a purely spatial distance $r$ so that $\sigma = \frac{1}{2}r^2$. Consulting e.g. \cite{PeskinSchroeder}  we find that in the Minkowski vacuum state,
\begin{equation}
\label{eq:auxGMink}
\mathcal{G}(x, x') = -\frac{1}{4\pi} \int \frac{dp}{\sqrt{p^2 + m^2}} e^{-i p r}, \\
\end{equation}
which can be related to the modified Bessel function of the first kind $K_0(\omega)$ via the identity
\begin{equation}
K_0(\omega) = \frac{1}{2} \int db \frac{e^{i\omega b}}{\sqrt{b^2 + 1}};
\end{equation}
thus
\begin{subequations}
\begin{align}
\mathcal{G}(x, x') &= \frac{1}{2\pi} K_0(m r) , \\
\label{eq:auxGexp}
& = -\frac{1}{4\pi} \left(\ln\left[ \left(\frac{1}{2}e^{2 \gamma_E}\right) \frac{1}{2} r^2\right] + \ln m^2 \right) + \mathcal{O}(r^2),
\end{align}
\end{subequations}
where $\gamma_E$ is the Euler-Mascheroni constant. Under the hypothesized conditions, the Hadamard form \eqref{eq:HadForm} is
\begin{equation}
\label{eq:HadFormFlat}
\mathcal{G}(x, x') = \alpha \left( V(x, x') \ln \left(\mu \frac{1}{2}r^2 + i \epsilon\right) + W(x, x') \right),
\end{equation}
which is consistent with \eqref{eq:auxGexp} given 
\begin{subequations}
\label{eq:QFTIdentifications}
\begin{align}
\mu &= \frac{1}{2} e^{2 \gamma_E}, \\
\alpha &= \frac{1}{4 \pi}, \\
\left[V(x, x')\right] &= \left[V_0(x, x')\right] = -1.
\end{align}
\end{subequations}

\bibliographystyle{apsrev4-1}
\bibliography{stresstensor}
\end{document}